\documentclass[11pt, a4paper]{article}
\usepackage[utf8]{inputenc}
\usepackage{amsmath}
\usepackage{amssymb}
\usepackage{graphicx}
\usepackage{geometry}
\usepackage[hidelinks]{hyperref}
\usepackage{authblk}
\usepackage{caption}
\usepackage{listings}
\usepackage{inconsolata}
\usepackage{xcolor}
\usepackage{siunitx}

\title{\textbf{ParamRF: A JAX-native Framework for\\Declarative Circuit Modelling}}
\author[1]{G.V.C. Allen}
\author[2]{D.I.L. de Villiers}
\affil[1, 2]{Stellenbosch University}
\date{\today}

\begin{document}

\definecolor{codegreen}{rgb}{0,0.6,0}
\definecolor{codegray}{rgb}{0.5,0.5,0.5}
\definecolor{codepurple}{rgb}{0.58,0,0.82}
\definecolor{backcolour}{rgb}{0.97,0.97,0.95}

\lstdefinestyle{mystyle}{
    backgroundcolor=\color{backcolour},   
    commentstyle=\color{codegreen},
    keywordstyle=\color{magenta},
    numberstyle=\tiny\color{codegray},
    stringstyle=\color{codepurple},
    basicstyle=\ttfamily\footnotesize,
    breakatwhitespace=false,         
    breaklines=true,                 
    captionpos=b,                    
    keepspaces=true,                 
    numbers=left,                    
    numbersep=5pt,                   
    showspaces=false,                
    showstringspaces=false,
    showtabs=false,                  
    tabsize=2
}
\lstset{style=mystyle}

\maketitle

\begin{abstract}
\noindent This work introduces \texttt{ParamRF}: a Python library for efficient, parametric modelling of radio frequency (RF) circuits. Built on top of the next-generation computational library JAX, as well as the object-oriented wrapper Equinox, the framework provides an easy-to-use, declarative modelling interface, without sacrificing performance. By representing circuits as JAX \textit{PyTrees} and leveraging just-in-time compilation, models are compiled as pure functions into an optimized, algebraic graph. Since the resultant functions are JAX-native, this allows computation on CPUs, GPUs, or TPUs, providing integration with a wide range of solvers. Further, thanks to JAX's automatic differentiation, gradients with respect to both frequency and circuit parameters can be calculated for any circuit model outputs. This allows for more efficient optimization, as well as exciting new analysis opportunities. We showcase \texttt{ParamRF}'s typical use-case of fitting a model to measured data via its built-in solvers, which include classical optimizers like L-BFGS and SLSQP, as well as modern Bayesian samplers via BlackJAX and PolyChord. The result is a flexible framework for frequency-domain circuit modelling, fitting and analysis.
\end{abstract}

\section{Introduction}
The use of radio frequency (RF) circuit models is central to many fields in engineering, such as high-frequency electronics, power systems, and semiconductor design. At the core of these approaches is the need to create parametric circuit models, which are then fit to measured data. While powerful, GUI-driven commercial tools exist, they often lack the programmatic interface needed to easily define custom models or error functions, and are unable to integrate with modern solvers or statistical frameworks. Script-based approaches, on the other hand, can incur a performance overhead resulting from dynamic memory allocation and interpreter context switching, and lack the power and flexibility a more general framework can provide.

The recent emergence of computational deep learning frameworks based on just-in-time (JIT) compilation and automatic differentiation (AD) has introduced new possibilities in scientific computing. Libraries such as JAX \cite{jax2018github} offer an integrated linear algebra solution with a high-level syntax that mirrors popular Python libraries such as NumPy. JAX differs from NumPy, however, in that it is able to trace and then compile any pure mathematical function into a computational graph that can subsequently be read by the XLA (Accelerated Linear Algebra) compiler. This results in any unnecessary overhead being almost entirely eliminated, while retaining a descriptive and readable syntax. Due to the flexibility of the XLA ecosystem, the resultant code can also be run on a number of different architectures, including modern GPUs and TPUs. Further, thanks to the algebraic representation, all functions become end-to-end differentiable, allowing full derivatives to be calculated.

This work introduces \texttt{ParamRF}: a Python library for declarative circuit modelling. Built on top of JAX, as well as the helper package Equinox \cite{kidger2021equinox} (which allows for the definition of JAX \textit{PyTrees} via Python objects), the library is designed to provide an easy-to-use, declarative, high-performance, frequency-domain circuit modelling interface. Since JAX leverages the XLA architecture, any definition of a circuit model's properties (such as its S- or ABCD-parameters) can be JIT-compiled and optimized. This is a major benefit over implementations that might be built using libraries such as NumPy or even lower-level languages. Since the mathematics of the model can be simplified and even vectorized using the compiler, performance is improved. This is especially important for complex, high-dimensional circuit models, where a naive implementation could result in the re-computation of sub-model properties. Further, thanks to JAX's automatic differentiation, gradients can be calculated with respect to both frequency and circuit parameters, which can be utilized for more efficient gradient-based optimization, as well as exciting new circuit analysis opportunities. The library's main goals are summarized as follows:
\begin{enumerate}
    \item Provide a simple, declarative, and object-oriented API for building complex circuit models.
    \item Offer an extensible, high-level fitting interface that can use a wide range of fitting techniques, including both classical optimizers and modern Bayesian inference libraries.
    \item Allow for differentiation of model outputs, both with respect to frequency and model parameters.
\end{enumerate}

This paper provides an overview of the library's features, and demonstrates its typical use of fitting a circuit model to measured data.

\section{Modelling}

The object-oriented architecture of \texttt{ParamRF} is based on the principles of the Equinox library, which represents models as both Python \textit{dataclasses} and JAX \textit{PyTrees}. Although a deep understanding of these concepts is not necessary, the reader is encouraged to briefly study the JAX, Equinox and Python documentations for further details. Ultimately, the representation of models as objects enables both a declarative and compositional circuit modelling approach: parameters and sub-models are listed as dataclass \textit{fields} in the initial class definition, and functions are defined as dataclass \textit{methods} that operate on those fields. This clear separation of attributes and functionality allows not only for self-documenting code, but also for the straightforward definition and composition of complex models.

\subsection{Class Hierarchy}
At the core of the \texttt{ParamRF} library are the \texttt{Model} and \texttt{Frequency} classes and a number of parameter wrappers, upon which all models are built. The \texttt{Model} class represents the base class for any computable RF component, which can include ``foundational'' models described using equations (resistors, transmission lines etc.) and ``composite'' circuit models. Users who wish to define their own custom models should therefore inherit from this class, using a combination of parameters and \texttt{Model} objects as the model attributes.

Note that, in contrast to the popular library scikit-rf \cite{scikit-rf} (which this library aims to complement and not replace), models are \textit{functional} in nature as opposed to \textit{data} driven. Specifically, models only contain their parameters and functional methods, in contrast to storing their S-parameter and frequency arrays directly. Following this, as well as the paradigm of JAX and Equinox, models are \textit{immutable}, meaning that parameters are ``frozen'' for a given instance. This aligns with the pure approach required by JAX, but may present a new perspective to some readers. Creating a new model with updated parameters, however, is still straightforward, using the higher-level method \texttt{Model.at}, or directly using Equinox's \texttt{tree\_at()} or dataclass's \texttt{replace()}.

Typically, foundational models only contain parameters, with the appropriate method being defined to compute one of their primary properties such as \texttt{s} or \texttt{a} as a function of frequency. Composite models, on the other hand, typically contain and compose other \texttt{Model} objects in a nested fashion, and typically override the Python \texttt{\_\_call\_\_} method to return the final composed model. There is, however, no restriction on the combination of these two approaches: any class method implemented by the user will be used instead of the library's built-in conversion utilities. 

\newpage
\subsection{Model Composition}
\texttt{ParamRF} provides several commonly-used models such as lumped and distributed elements, as well as convenience models that implement functionality like cascading and port termination. These can be used to directly build other models in a compositional approach. For example, the Python ** operator can be used for two-port cascading (drawing inspiration from scikit-rf). This is demonstrated in Listing \ref{lst:composition}, where an RLC filter is created. Note that the resultant \texttt{rlc} is a first-class \texttt{Model} of type \texttt{Cascade}, consisting of $R$, $L$ and $C$ parameters.

\begin{lstlisting}[language=Python, caption={Model composition in ParamRF using operator overloading.}, label={lst:composition}]
from pmrf.models import Resistor, Inductor, ShuntCapacitor
from pmrf.parameters import Fixed, Bounded

# Instantiate the lumped element models. L and C can be varied, whereas R is fixed.
resistor = Resistor(R=Fixed(1000.0))
inductor = Inductor(L=Bounded(1.0, 100.0, scale=1e-9))
capacitor = ShuntCapacitor(C=Bounded(1.0, 100.0, scale=1e-12))

# Cascade the models, storing the result
rlc = resistor ** inductor ** capacitor
\end{lstlisting}

\subsection{Model Definition}
A key goal of \texttt{ParamRF} is to reduce boilerplate and make custom model definitions clear and self-documenting. Any attributes can therefore be added as fields to a model class, however attributes are classified as either ``static'' or ``dynamic'' in nature. Specifically, built-in Python types such as \texttt{str}, \texttt{int}, \texttt{list} etc. are seen as static in the model hierarchy, whereas parameters and other models are dynamic and can be adjusted by fitting routines. Initialization can also be made flexible via the \texttt{param} field specifier, allowing parameters to be populated directly with a float value, a JAX array, or using factory methods such as \texttt{Bounded}, \texttt{Random} or \texttt{Fixed}. The factory methods are the most powerful, allowing for the definition of a scale, bounds, a prior probability distribution, and more. Listing \ref{lst:declarative} provides an example demonstrating some of these concepts. Note how the \texttt{s} function accepts frequency as input - the model does \textit{not} store its frequency.

\begin{lstlisting}[language=Python, caption={Definition of an amplifier terminated in a non-ideal resistor model, with the gain bounded between $10$ and $15$ dB, and resistance fixed to $\SI{1}{k\ohm}$. Parameters and sub-models are defined as class attributes. JAX's numpy interface is utilized for calculations.}, label={lst:declarative}]
import jax.numpy as jnp
import pmrf as prf
from pmrf.models import Resistor, PiCLC, SModel
from pmrf.parameters import Bounded, Fixed

# Define a model class. Behaviour is defined by implementing 
# a primary matrix function, such as "s" as below.
class TerminatedAmplifier(prf.Model):
    gain: prf.Param = Bounded(prf.math.db_2_mag(10), prf.math.db_2_mag(15))
    resistor: prf.Model = Resistor(Fixed(1.0, scale=1e3))
    parasitics: prf.Model = PiCLC(C1=0.05e-12, L=0.1e-9, C2=0.1e-12)

    def s(self, freq: prf.Frequency) -> jnp.ndarray:       
        # We use jnp for calculations as a function of freq.f or freq.w.
        # We also create a fixed "SModel" for the gain matrix for easy cascading.
        # Note that "terminated()" defaults to a SHORT,
        # and that SModel only requires its frequency for interpolation
        s21 = jnp.sqrt(self.gain) * jnp.ones_like(freq.f)
        zeros = jnp.zeros_like(freq.f)
        amp = SModel(freq, jnp.array([
            [zeros, zeros],
            [s21,   zeros]
        ]).transpose(2, 0, 1))
        return (amp ** self.parasitics ** self.resistor).terminated().s(freq)
\end{lstlisting}

\section{Evaluation and Differentiation}
Similar to scikit-rf, once a user defines a model, a number of attributes are automatically generated, for example \texttt{s\_db}, \texttt{s\_mag} etc. Because each of these are pure functions (given the user-defined function is pure), the entire system is compatible with \texttt{jax.grad}, as well as Equinox \textit{partitioning}. Further, since the model is a JAX \texttt{PyTree}, a ``gradient'' model can be created with respect to a chosen function, which has parameter values equal to the evaluated derivative for that function. This is demonstrated in Listing \ref{lst:autodiff}, where the model is differentiated with respect to an arbitrary, custom function.

\begin{lstlisting}[language=Python, caption={A demonstration of automatic differentiation of a custom model function with respect to a desired parameter, using Equinox partitioning and JAX grad.}, label={lst:autodiff}]
import jax.numpy as jnp
import equinox as eqx
import pmrf as prf

# Define a function that returns a scalar value for the S11 magnitude
# with respect to a constant 0.1 and using explicit equinox partitioning
def custom_fn(params, static, freq):
    model = eqx.combine(params, static)
    s11_mag = model.s_mag(freq)[:, 0, 0]
    return jnp.mean((s11_mag - 0.1) ** 2)

# Setup the model and frequency
freq = prf.Frequency(1, 10, 9, 'GHz')
model = TerminatedAmplifier()

# Compute the gradient of the loss with respect to all parameters
params, static = eqx.partition(model, eqx.is_inexact_array)
grad_model = jax.grad(custom_fn)(params, static, freq)

# The gradient is a PyTree with the same structure as the model
C1_grad = grad_model.parasitics.C1
\end{lstlisting}

\section{Fitting, Optimization, and Inference}
\texttt{ParamRF} provides high-level \texttt{fitting}, \texttt{optimize} and \texttt{infer} sub-modules designed to abstract the boilerplate of interfacing with different solver backends. The fitting module supports both frequentist and Bayesian techniques via solver routing. Although the solver interface is extensible (solvers can be extended by deriving from the relevant abstract interfaces in \texttt{optimize} and \texttt{infer}), a number of built-in fitters are available.

\paragraph{Frequentist Optimization:} For general frequentist-based fitting, \texttt{ParamRF} provides wrappers around SciPy's ``minimize'' \cite{2020SciPy-NMeth} (in \texttt{ScipyMinimize}) and Optimistix  \cite{optimistix2024} (in \texttt{OptimistixMinimise}). This provides access to algorithms such as SLSQP, Nelder-Mead, and L-BFGS, amongst others. The use of the JAX-native Optimistix allows the entire optimization loop to be JIT-compiled.

\paragraph{Bayesian Inference:} For statistical inference and model comparison, a number of modern Bayesian samplers are available. Support for MCMC methods via BlackJAX \cite{cabezas2024blackjax} is included, providing access to algorithms such as the No-U-Turn Sampler (NUTS). Support for nested sampling approaches is also available from PolyChord \cite{handley2015polychord}. These backends ultimately allow for high-dimensional parameter estimation and inference, as well as Bayesian evidence calculation for circuit model comparison.

\newpage
\section{Example Application: Fitting a Model}

To demonstrate the library's usage, we show a complete example of fitting the built-in \texttt{CoaxialLine} model to the measurement of a $\SI{10}{m}$ coaxial cable (provided as an example in the repository). Data is loaded using scikit-rf; the model is instantiated with appropriate initial parameters; the solver is configured; fitting is run; and results are plotted using scikit-rf.

\begin{lstlisting}[language=Python, caption={A typical fitting workflow using the \texttt{ScipyMinimize} solver.}, label={lst:fitting}]
import pmrf as prf
import skrf as rf

from pmrf.models import CoaxialLine
from pmrf.parameters import Bounded, Fixed
from pmrf.optimize import ScipyMinimize
from pmrf.fitting import fit_minimize

# Load the measured data and setup the model
measured = rf.Network('data/10m_cable.s2p', f_unit='MHz')
model = CoaxialLine(
    din = Bounded(1.0, 1.24, scale=1e-3),
    dout = Bounded(3.0, 3.4, scale=1e-3),
    epr = Bounded(1.4, 1.5),
    rho = Bounded(1.4, 1.8, scale=1e-8),
    tand = Bounded(0.0, 0.01, value=0.0, scale=0.01),
    mur = Fixed(1.0),
    length = Bounded(9.0, 11.0),
)

# Initialize the solver.
solver = ScipyMinimize(method='Nelder-Mead')

# Run the fit. We fit on the real and imaginary "features"
results = fit_minimize(
    model=model,
    data=measured,
    solver=solver,
    features=['s11_re', 's11_im'],
)

# Plot the resultant S11
freq = prf.Frequency.from_skrf(measured.frequency)
results.model.plot_s_db(freq, m=0, n=0)
measured.plot_s_db(m=0, n=0)

\end{lstlisting}

\section{Position within the Ecosystem}
\texttt{ParamRF} is designed to complement, not replace, existing tools. It integrates with the scientific Python ecosystem by leveraging a large range of open-source tools:
\begin{itemize}
    \item \textbf{JAX and Equinox:} Provides the foundation for high-performance, differentiable computation, as well as declarative modelling.
    \item \textbf{scikit-rf:} Used as an interface for handling and plotting measured data and fitted models, as well as a reference for some of the framework's design and mathematical functions.
    \item \textbf{SciPy and Optimistix:} Serve as backends for frequentist optimization techniques.
    \item \textbf{BlackJAX and PolyChord:} Serve as backends for Bayesian sampling approaches, providing MCMC and nested sampling algorithms.
\end{itemize}

\section{Conclusion}
\texttt{ParamRF} provides a declarative API for circuit modelling, while building on top of the next-generation computing library JAX. By representing RF circuits as composable PyTrees via Equinox, it provides a flexible tool for circuit modelling, fitting and analysis. Future work will focus on expanding the library of built-in models and fitters; adding support for more complex circuit topologies (such as parallel and hybrid connections); and providing built-in tools for inference and analysis.

The code and documentation can be found on GitHub at https://github.com/gvcallen/paramrf. Contributions are always welcome.

\section*{Acknowledgements}
We thank the developers of JAX, Equinox, and the many open-source scientific libraries that made this work possible.

\bibliographystyle{IEEEtran} 
\bibliography{main}

@software{jax2018github,
  author = {James Bradbury and Roy Frostig and Peter Hawkins and Matthew James Johnson and Chris Leary and Dougal Maclaurin and George Necula and Adam Paszke and Jake Vander{P}las and Skye Wanderman-{M}ilne and Qiao Zhang},
  title = {{JAX}: composable transformations of {P}ython+{N}um{P}y programs},
  url = {http://github.com/jax-ml/jax},
  version = {0.3.13},
  year = {2018},
}

@article{kidger2021equinox,
    author={Patrick Kidger and Cristian Garcia},
    title={{E}quinox: neural networks in {JAX} via callable {P}y{T}rees and filtered transformations},
    year={2021},
    journal={Differentiable Programming Workshop at Neural Information Processing Systems}
}

@ARTICLE{scikit-rf,
author={Arsenovic, Alexander and Hillairet, Julien and Anderson, Jackson and Forst\'en, Henrik and Rie{\ss}, Vincent and Eller, Michael and Sauber, Noah and Weikle, Robert and Barnhart, William and Forstmayr, Franz},
journal={IEEE Microwave Magazine},
title={scikit-rf: An Open Source Python Package for Microwave Network Creation, Analysis, and Calibration [Speaker's Corner]},
year={2022},
volume={23},
number={1},
pages={98-105},
doi={10.1109/MMM.2021.3117139}}

@article{handley2015polychord,
  title={{POLYCHORD}: next-generation nested sampling},
  author={Handley, WJ and Hobson, MP and Lasenby, AN},
  journal={Monthly Notices of the Royal Astronomical Society},
  volume={453},
  number={4},
  pages={4384--4398},
  year={2015},
  publisher={Oxford University Press}
}

@misc{cabezas2024blackjax,
      title={{B}lack{JAX}: {C}omposable {B}ayesian inference in {JAX}},
      author={Alberto Cabezas and Adrien Corenflos and Junpeng Lao and Rémi Louf},
      year={2024},
      eprint={2402.10797},
      archivePrefix={arXiv},
      primaryClass={cs.MS}
}

@article{optimistix2024,
    title={Optimistix: modular optimisation in JAX and Equinox},
    author={Jason Rader and Terry Lyons and Patrick Kidger},
    journal={arXiv:2402.09983},
    year={2024},
}

@ARTICLE{2020SciPy-NMeth,
  author  = {Virtanen, Pauli and Gommers, Ralf and Oliphant, Travis E. and
            Haberland, Matt and Reddy, Tyler and Cournapeau, David and
            Burovski, Evgeni and Peterson, Pearu and Weckesser, Warren and
            Bright, Jonathan and {van der Walt}, St{\'e}fan J. and
            Brett, Matthew and Wilson, Joshua and Millman, K. Jarrod and
            Mayorov, Nikolay and Nelson, Andrew R. J. and Jones, Eric and
            Kern, Robert and Larson, Eric and Carey, C J and
            Polat, {\.I}lhan and Feng, Yu and Moore, Eric W. and
            {VanderPlas}, Jake and Laxalde, Denis and Perktold, Josef and
            Cimrman, Robert and Henriksen, Ian and Quintero, E. A. and
            Harris, Charles R. and Archibald, Anne M. and
            Ribeiro, Ant{\^o}nio H. and Pedregosa, Fabian and
            {van Mulbregt}, Paul and {SciPy 1.0 Contributors}},
  title   = {{{SciPy} 1.0: Fundamental Algorithms for Scientific
            Computing in Python}},
  journal = {Nature Methods},
  year    = {2020},
  volume  = {17},
  pages   = {261--272},
  adsurl  = {https://rdcu.be/b08Wh},
  doi     = {10.1038/s41592-019-0686-2},
}

\end{document}